\documentclass[twocolumn,fp]{jpsj3}

\usepackage{txfonts}

\usepackage{amsmath}
\usepackage{amsfonts}
\usepackage{bm}
\usepackage{graphicx}
\usepackage{color}
\usepackage{algorithm,algorithmic}

\title{
Creation and manipulation of quantized vortices in Bose-Einstein condensates
using reinforcement learning
}

\author{Hiroki Saito}
\inst{Department of Engineering Science, University of
Electro-Communications, Tokyo 182-8585, Japan}

\abst{
We apply the technique of reinforcement learning to the control of nonlinear
matter waves.
In this method, an agent controls the position, strength, and shape of an
external Gaussian potential to create and manipulate quantized vortices in a
Bose-Einstein condensate (BEC) trapped in a harmonic potential.
The density and velocity distributions of the BEC at each moment obtained by
the Gross-Pitaevskii evolution are directly input into a convolutional
neural network to determine the next action of the agent.
We demonstrate that a stationary single-vortex state can be produced in a
two-dimensional system, and a stationary vortex-ring state can be produced
in a three-dimensional system.
}

\begin{document}
\maketitle

\section{Introduction}

Recent developments in machine learning have been remarkable.
A standout example is the computer program
``AlphaGo''~\cite{Silver16,Silver17}, which defeated the best human players
in the game of Go.
In AlphaGo, reinforcement learning with deep neural networks (called deep-Q
learning) is used to evaluate the situation of the game and determine the
next action.
Another interesting example of the use of deep-Q learning was reported in
Ref.~\citen{Mnih}, in which a computer agent playing video games is trained
to get higher scores.
It was demonstrated that the computer agent outperforms human players after
training, without prior knowledge about the games.

In this study, we focus on the control of quantum systems by reinforcement
learning~\cite{Chen, Bukov, BukovB, Arria, Yu, Fosel, Zhang, Andreasson,
  Chen19, August, Niu, JChen, Wang, Macke}.
Controlling quantum systems and producing desired quantum states are
important in a variety of areas in quantum physics.
Reinforcement learning consists of an agent and the environment.
In this case, the quantum system is regarded as the environment.
An easily accessible initial state, such as the ground state, is first
prepared, and during the time evolution, the agent makes decisions to
control the time-dependent parameters in the Hamiltonian.
The quantum state develops depending on the parameters determined by the
agent, and the information of the quantum state is fed back to the agent.
Depending on the quantum state, a reward is also given to the agent,
and the agent is trained to maximize the reward.
The reward is, for example, the overlap or fidelity between the controlling
state and the target state, 

In this paper, we consider a Bose-Einstein condensate (BEC) of an atomic gas
as a quantum system to be controlled.
Optimal control of a BEC has been studied for a variety of purposes,
such as spatial transport~\cite{Hohenester, XChen, Amri},
splitting into two BECs~\cite{Hohenester,Jager,Menne},
squeezing of quantum fluctuations~\cite{Grond, Jager15},
excitation to specific states~\cite{Bucker, Jager, Frank, Hocker, Weidner,
Sorensen},
changing trap geometry~\cite{Menne,Riahi},
improving interferometry~\cite{Saywell},
and creating a self-bound droplet~\cite{Menne19}.
In these studies, quantum control theories, such as GRAPE~\cite{Khaneja} and
CRAB~\cite{Doria, Caneva} are used.
Recently, machine learning techniques have been used for the optimized
creation of a BEC~\cite{Wigley,Nakamura,Barker,Davletov} as well as the
transport and decompression~\cite{Henson} of a BEC.

\begin{figure}[tb]
\begin{center}
  \includegraphics[width=5cm]{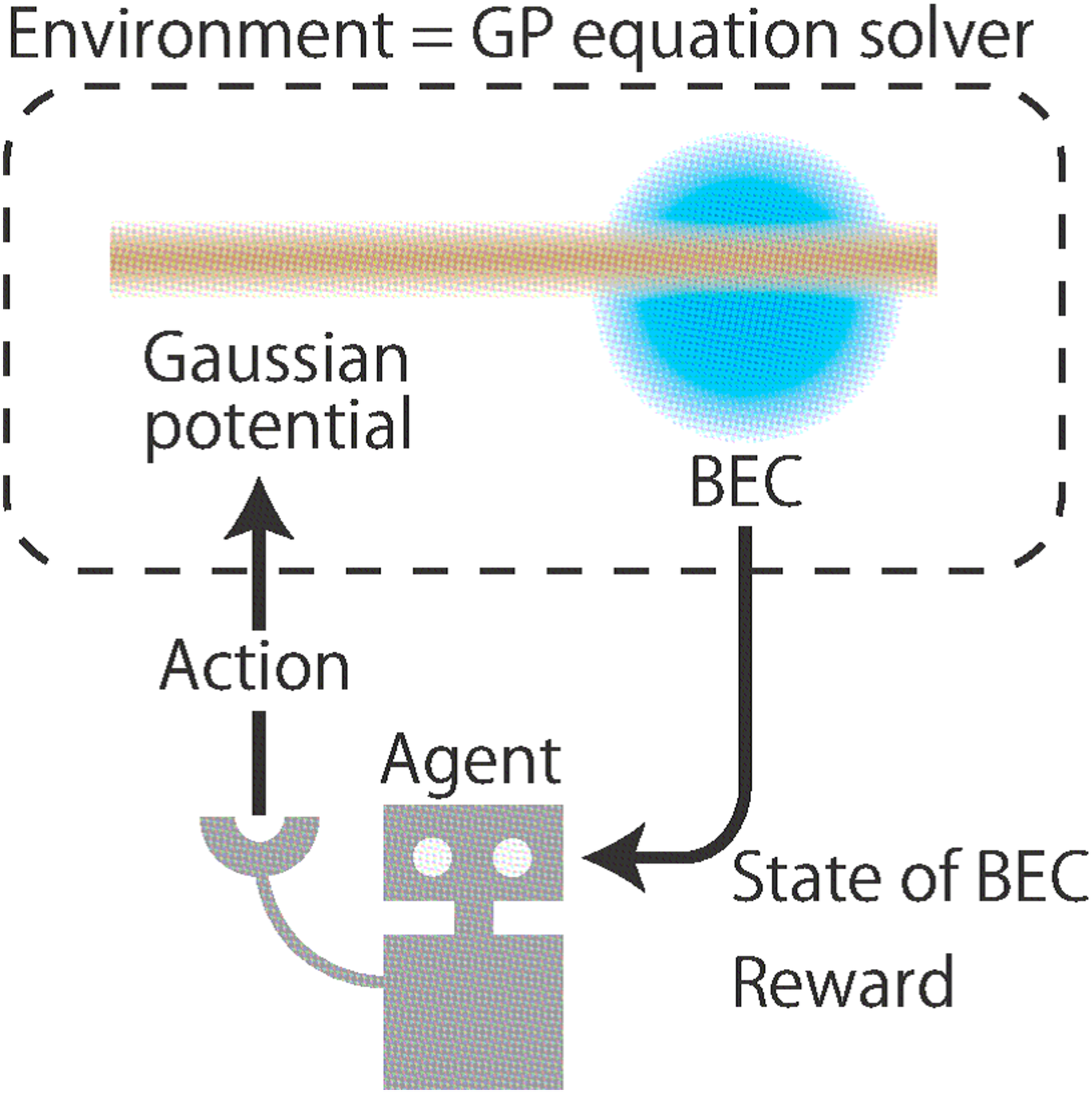}
\end{center}
\caption{
  Schematic illustration of our system.
  The system consists of an agent and the environment.
  At each time step, the agent decides on an action on the environment,
  namely, how to change the external Gaussian potential applied to the BEC.
  The dynamics of the BEC with the time-dependent Gaussian potential is
  obtained by numerically solving the Gross-Pitaevskii (GP) equation, and
  the state of the BEC and the reward are given to the agent.
  The agent is trained to get a higher reward, which leads to the production
  of a desired state of the BEC.
}
\label{f:schematic}
\end{figure}
Figure~\ref{f:schematic} illustrates reinforcement learning to control a
BEC.
We aim to create and manipulate quantized vortices in a BEC by changing a
Gaussian external potential.
Experimentally, such a potential can be produced by a nonresonant Gaussian
laser beam applied to the BEC.
On the environmental side, the dynamics of the BEC with a time-dependent
Gaussian external potential are obtained by numerically solving the
Gross-Pitaevskii (GP) equation.
At each time step, the agent obtains the state of the BEC from the
environment, and decides on the next action, i.e., how to change the
position, strength, and shape of the Gaussian potential.
The decision of the agent is made using a deep convolutional neural network
(CNN), where the spatial distributions of the density and velocity of the
BEC are directly input into the CNN, giving the next action as an output.
At each time step, the agent also receives reward from the environment.
The reward is higher when the state of the BEC is closer to the target
state.
The CNN of the agent is trained to maximize the total amount of rewards,
which enables the agent to discover an optimal control of the Gaussian
potential.

Using the above scheme, we demonstrate that quantized
vortices can be created and manipulated in two-dimensional (2D) and
three-dimensional (3D) BECs.
In a 2D system, the target state is set to be the state with a
singly-quantized vortex at the center.
Although both vortices and antivortices are created in pairs by a simple
translation of the Gaussian potential~\cite{Frisch,Neely}, the agent finds a
clever way to generate a single vortex state.
In a 3D system, the target state is set to be the stationary vortex-ring
state, and the agent finds that it can be created by a simple Gaussian
laser beam.

The remainder of the paper is organized as follows.
Section~\ref{s:method} explains our method, Sec.~\ref{s:results} shows the
numerical results, and Sec.~\ref{s:conc} gives the conclusions to this
study.

\section{Method}
\label{s:method}

\subsection{Environment} \label{s:env}

We begin by describing the environment of the reinforcement learning
system.
As described above, the environment receives the action from the agent and
evolves to the next time step depending on the action.
The environment then relays its state to the agent.
The environment also evaluates whether the action of the agent was a good
action, and gives a reward to the agent.

As an environment, we consider a BEC of bosonic atoms with mass $m$ confined
in a harmonic potential.
In the mean-field approximation at zero temperature, the macroscopic wave
function $\psi(\bm{r}, t)$ obeys the GP equation given by
\begin{equation} \label{GP}
i\hbar \frac{\partial \psi}{\partial t} = -\frac{\hbar^2}{2m} \nabla^2 \psi
+ V_{\rm trap}(\bm{r}) \psi + V_G(\bm{r}, t) \psi + g |\psi|^2 \psi,
\end{equation}
where $V_{\rm trap} = m [\omega_\perp^2 (x^2 + y^2) + \omega_z^2 z^2] / 2$
is the harmonic trap potential with $\omega_\perp$ and $\omega_z$ being the
radial and axial trap frequencies, $V_G$ is the Gaussian external potential,
and $g = 4 \pi \hbar^2 a / m$ is the interaction coefficient with $a$ being
the $s$-wave scattering length of the atom.
The wave function is normalized as $\int |\psi|^2 d\bm{r} = N_{\rm atom}$,
where $N_{\rm atom}$ is the number of atoms.

In the next section, we will investigate both 2D and 3D systems.
For the 3D system, for simplicity, the harmonic trap potential is assumed to
be isotropic, i.e., $\omega_\perp = \omega_z \equiv \omega$.
The external Gaussian potential for the 3D problem has the form
\begin{equation} \label{3DGauss}
  V_G(\bm{r}, t) = A(t) \exp \left\{ \frac{y^2}{d_y^2(t)}
    + \frac{[z - \zeta(t)]^2}{d_z^2} \right\},
\end{equation}
where the strength $A(t) > 0$, width $d_y(t) > 0$, and position $\zeta(t)$
are control parameters, and the width $d_z$ is a constant.
Such an external potential can be produced by a blue-detuned Gaussian laser
beam propagating in the $x$ direction.

The 2D system is experimentally realized by tight confinement in the $z$
direction such that $\hbar \omega_z$ is much larger than the other
characteristic energies.
In this case, the wave function can be approximated as $\psi(\bm{r}, t) =
\psi_\perp(x, y, t) \psi_0(z) e^{-i \omega_z t / 2}$, where $\psi_0(z)$ is
the ground state of the harmonic oscillator potential $m \omega_z^2 z^2 /
2$.
Multiplying Eq.~(\ref{GP}) by $\psi_0(z)$ and integrating it with respect to
$z$, the system is reduced to 2D as
\begin{equation} \label{GP2D}
i\hbar \frac{\partial \psi_\perp}{\partial t} = -\frac{\hbar^2}{2m}
\nabla_\perp^2 \psi_\perp + V_{\rm trap}(x, y) \psi_\perp + V_G(x, y , t)
\psi_\perp + g_\perp |\psi_\perp|^2 \psi_\perp,
\end{equation}
where $\nabla_\perp^2 = \partial^2 / \partial x^2 + \partial^2 / \partial
y^2$, $V_{\rm trap}(x, y) = m \omega_\perp^2 (x^2 + y^2) / 2$, and $g_\perp
= [m \omega_z / (2\pi\hbar)]^{1/2} g$.
The external Gaussian potential for the 2D problem is assumed to be
\begin{equation} \label{2DGauss}
  V_G(x, y, t) = A_\perp(t) \exp\left\{ \frac{[x - \xi(t)]^2}{d^2}
  + \frac{[y - \eta(t)]^2}{d^2} \right\},
\end{equation}
where the strength $A_\perp(t) > 0$ and position $\xi(t)$, $\eta(t)$ are
control parameters, and the width $d$ is a constant.
This form of the potential is generated by a blue-detuned Gaussian laser
beam propagating in the $z$ direction.

At the start of each run of time evolution (called an episode), the
environment is reset: the control parameters are set to the initial values,
and the wave function is set to the ground state for $V_G(t = 0)$.
The ground-state wave function is prepared by the imaginary-time evolution,
where $i$ on the right-hand sides of Eqs.~(\ref{GP}) and (\ref{GP2D}) is
replaced with $-1$.
The real-time and imaginary-time evolution is numerically obtained by the
pseudospectral method~\cite{recipe}.
After the initial reset of the environment, the real-time evolution starts.
At intervals of time $\Delta t$ (which is much larger than the discretized
time interval $\delta t$ in the pseudospectral method), the environment
returns its state and the reward to the agent, and receives the action from
the agent (see Fig.~\ref{f:schematic}).
The state of the environment given to the agent is the density distribution
$\rho = |\psi|^2$ and flux distribution $\bm{F} = \hbar (\psi^* \bm{\nabla}
\psi - \psi \bm{\nabla} \psi^*) / (2mi)$ ($\psi \rightarrow \psi_\perp$ for
2D) of the BEC.
The state of the environment also includes the current shape of the Gaussian
potential $V_G$ so that the agent can determine how to change the
potential.
The reward given to the agent is calculated from the overlap between the
current wave function and the target wave function, which is specified in
the next section.
By the action received from the agent, one of the parameters in the Gaussian
external potential is changed by a small amount at each time step.
Each episode terminates at $t = 500 \Delta t \equiv T_{\rm end}$, which is
followed by the next episode.

\subsection{Agent}

The agent observes the state $s_t$ of the environment at time $t$, and
determines the action $a_t$ on the environment.
The agent then gets the reward $r_t$ for the action from the environment.
From these experiences, the agent tries to maximize the discounted
cumulative reward, $\sum_{n=0}^\infty \gamma^n r_{t+n \Delta t}$,
where $0 < \gamma < 1$ is the discount rate~\cite{textbook2}.

\begin{figure}[tb]
\begin{center}
  \includegraphics[width=8cm]{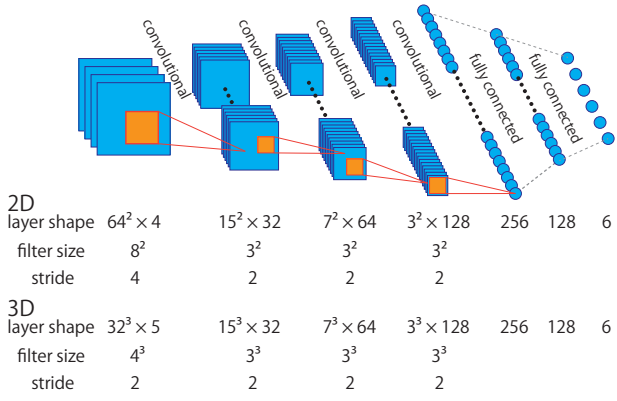}
\end{center}
\caption{
  Structures of the CNNs for the 2D and 3D cases.
  The density and flux distributions of the BEC and the Gaussian potential
  distribution are input into the CNN.
  The six values from the output correspond to $Q(s_t, a)$ for six actions
  $a = 0, 1, \cdots, 5$.
  Filters are applied with the indicated stride and without padding in each
  convolutional layer.
  There are no pooling layers.
}
\label{f:cnn}
\end{figure}
As mentioned in Sec.~\ref{s:env}, as the state $s_t$ of the environment, we
take the density distribution $\rho(\bm{r}, t)$, flux distribution
$\bm{F}(\bm{r}, t)$, and the Gaussian potential $V_G(\bm{r}, t)$.
For the 2D case, the four distributions $\rho(x, y)$, $F_x(x, y)$, $F_y(x,
y)$, and $V_G(x, y)$ are expressed by $64 \times 64 \times 4$ pixels, and
are input into the CNN~\cite{textbook} shown in Fig.~\ref{f:cnn}.
For the 3D case, the five distributions $\rho(x, y, z)$, $F_x(x, y, z)$,
$F_y(x, y, z)$, $F_z(x, y, z)$, and $V_G(x, y, z)$ are expressed by $32
\times 32 \times 32 \times 5$ pixels.
The CNN consists of four convolutional layers and two fully-connected
layers.
The CNN outputs six values, which we denote $Q(s_t, a)$ with $a = 0, 1,
\cdots, 5$.
The agent determines the action $a_t$ by the $\epsilon$-greedy policy:
\begin{equation}
  a_t = \left\{ \begin{array}{cc}
    \mbox{randomly chosen from 0 to 5} & (r < \epsilon) \\
    \mbox{argmax}_a Q(s_t, a) & (r \geq \epsilon),
  \end{array} \right.
\end{equation}
where $0 \leq r < 1$ is a random number and $\mbox{argmax}_a Q(s_t, a)$
indicates the value of $a$ that maximizes $Q(s_t, a)$.
The value of $\epsilon$ is linearly decreased from 1 to 0.1 in the first
50,000 steps (100 episodes), and a constant value of $\epsilon = 0.1$ is
used after that.
The action $a_t$ chosen by the agent is applied to the environment, where
the Gaussian external potential is changed depending on $a_t$, and the BEC
evolves from $t$ to $t + \Delta t$.
The agent then receives the reward $r_t$ from the environment, and observes
the new state $s_{t + \Delta t}$.

According to the Bellman optimality equation~\cite{textbook2}, $Q(s_t, a_t)$
should approach the target value $r_t + \gamma Q(s_{t+\Delta t}, a')$,
where $a' = \mbox{argmax}_a Q(s_{t+\Delta t}, a)$.
It was reported in Ref.~\citen{Hasselt} that the training process is
stabilized by replacing $Q$ with $\hat Q$ in the above target value,
where $\hat Q$ is generated by another CNN, called the target network.
Thus, the target value is given by
\begin{equation} \label{yt}
  y_t = r_t + \gamma \hat Q(s_{t+\Delta t}, \mbox{argmax}_a
  Q(s_{t+\Delta t}, a)).
\end{equation}
The original CNN to generate $Q$ is updated in every training step, and the
original CNN is copied to the target CNN every $C$ steps, where we take $C
= 1000$.
The target value $y_t$ is thus calculated by fixing the target CNN during
$C$ steps, which stabilizes the learning process.
This method with Eq.~(\ref{yt}) is called double-Q
learning~\cite{Hasselt}.

The agent (CNN) is trained as follows.
In every time step, the set of data $(s_t, a_t, r_t, s_{t + \Delta t})$
is stored in the memory, which is called the replay memory~\cite{Mnih}.
In the present case, the replay memory can store 50,000 sets of data.
Once the limit is reached, old data are removed to store new data (queue).
From the replay memory, 32 sets of data are taken randomly, which is called
a minibatch.
For each dataset in the minibatch, the target value $y_t$ in Eq.~(\ref{yt})
is calculated, and the network parameters of the CNN are updated to minimize
$\sum_{\rm minibatch} L(Q(s_t, a_t) - y_t)$, where the function $L$ is
called the Huber loss~\cite{Huber}, defined by
\begin{equation}
  L(x) = \left\{ \begin{array}{cc}
    \frac{x^2}{2} & (|x| \leq 1) \\
    |x| - \frac{1}{2} & (|x| > 1).
    \end{array} \right.
\end{equation}
Using the Huber loss, large gradients are suppressed and network updates are
stabilized.
Updating the CNN is performed using the Adam optimization
scheme~\cite{Kingma} with a learning rate of $10^{-5}$-$10^{-4}$.
The procedure for the agent is summarized in Algorithm 1.

\begin{algorithm} 
\caption{}  
\label{algo}  
\begin{algorithmic}
\STATE Initialize deep-Q network $Q$
\STATE Initialize target network as $\hat Q = Q$
\FOR{episode = 1, $N_{\rm episode}$}
\STATE Initialize environment and get initial observation $s_0$
\FOR{$t=0$, $T_{\rm end}$} 
\STATE Select action $a_t = {\rm argmax}_a Q(s_t, a)$ with $\epsilon$-greedy
exploration
\STATE Execute action $a_t$ on environment and get reward $r_t$ and next
state $s_{t+\Delta t}$
\STATE Store $(s_t, a_t, r_t, s_{t+\Delta t})$ in replay memory
\STATE Sample minibatch of $(s_t, a_t, r_t, s_{t+\Delta t})$ from replay
memory
\STATE Set $y_t = r_t$ if $t = T_{\rm end}$, otherwise $y_t = r_t + \gamma
\hat Q(s_{t+\Delta t}, a')$, where $a' = {\rm argmax}_a Q(s_{t+\Delta t},
a)$
\STATE Train network using gradient of $L(y_t - Q(s_t, a_t))$
\STATE Copy $\hat Q = Q$ every $C$ steps
\ENDFOR 
\ENDFOR 
\end{algorithmic} 
\end{algorithm}

\section{Results}
\label{s:results}

We numerically demonstrate that our method can create and manipulate
vortices in BECs.
In the following, we normalize the length, time, and energy by $[\hbar / (m
\omega_\perp)]^{1/2}$, $\omega_\perp^{-1}$, and $\hbar \omega_\perp$,
respectively.
The density $|\psi_\perp|^2$ is normalized by $N_{\rm atom} m \omega_\perp /
\hbar$ in 2D, and $|\psi|^2$ is normalized by $N_{\rm atom} (m \omega_\perp
/ \hbar)^{3/2}$ in 3D.
For 2D calculations, the normalized interaction coefficient $\tilde g_\perp
= N_{\rm atom} a (8\pi m \omega_z / \hbar)^{1/2}$ is taken to be 1000.
For example, for $^{87}{\rm Rb}$ atoms in a quasi-2D trap with $\omega_\perp
= 2\pi \times 100$ Hz and $\omega_z = 2\pi \times 10$ kHz, $\tilde g_\perp =
1000$ corresponds to $N_{\rm atom} \simeq 4.1 \times 10^3$.
For 3D calculations, the normalized interaction coefficient $\tilde g =
4 \pi N_{\rm atom} a (m \omega_\perp / \hbar)^{1/2}$ is taken to be 6000.
For a 3D isotropic trap with $\omega = 2\pi \times 100$ Hz, $\tilde g =
6000$ corresponds to $N_{\rm atom} = 9.7 \times 10^4$.
The spatial discretization size is 0.15 with a $128 \times 128$ mesh for the
2D case, and 0.25 with a $64 \times 64 \times 64$ mesh for the 3D case.
These spatial data are averaged and reduced to $64 \times 64$ for the 2D
case and $32 \times 32 \times 32$ for the 3D case to be input into the CNN
in Fig.~\ref{f:cnn}.
The time interval for numerical integration is $\delta t = 0.002$, and the
time step for the reinforcement learning is $\Delta t = 50 \delta t = 0.1$.

\subsection{Two-dimensional system}

\begin{figure}[tb]
\begin{center}
  \includegraphics[width=8cm]{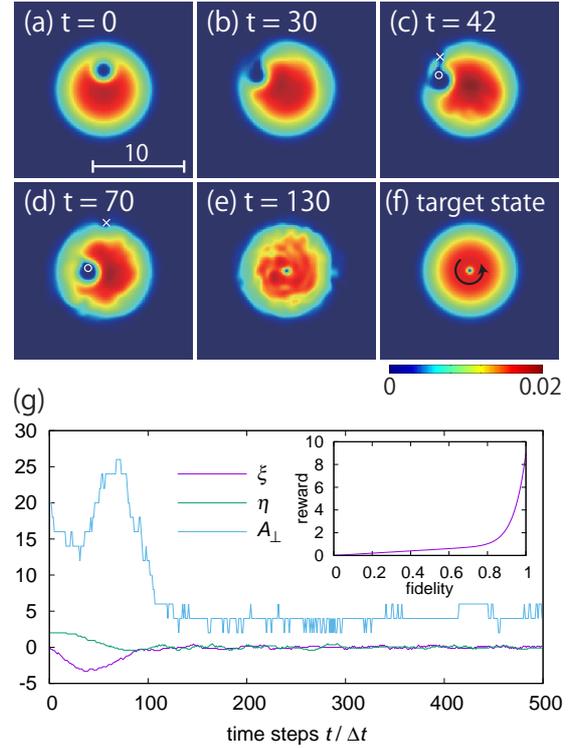}
\end{center}
\caption{
  Result with the best total reward through 5000 trials of episodes.
  (a)-(e) Snap shots of the density profiles in the time evolution.
  The crosses and circles in (c) and (d) indicate the positions of clockwise
  and counterclockwise vortex cores, respectively.
  (f) Density profile of the target state, which has a counterclockwise
  vortex at the center.
  (g) Parameters $\xi(t)$, $\eta(t)$, and $A_\perp(t)$ of the Gaussian
  external potential.
  The inset in (g) plots Eq.~(\ref{reward16}).
  See the Supplemental Material for a movie of the dynamics shown in
  (a)-(e)~\cite{movie2D}.
}
\label{f:ev2d}
\end{figure}

We first consider the 2D system, which obeys the GP equation in
Eq.~(\ref{GP2D}).
The initial parameters of the Gaussian potential in each episode are $\xi(0)
= 0$, $\eta(0) = 2$, and $A_\perp(0) = 20$.
The initial state of the BEC is the ground state for these parameters, as
shown in Fig.~\ref{f:ev2d}(a), where the density dip is due to the Gaussian
potential.
For this initial state, therefore, the rotational symmetry of the problem is
broken.
Here we set our target state $\psi_{\rm target}(x, y)$ to the lowest-energy
stationary state having a singly-quantized counterclockwise vortex at the
center, as shown in Fig.~\ref{f:ev2d}(f).
Numerically, such a wave function can be obtained by phase imprinting
followed by imaginary-time evolution.
Starting from the initial state without a vortex, the agent tries to produce
$\psi_{\rm target}(x, y)$ by controlling the position and strength of the
Gaussian potential in Eq.~(\ref{2DGauss}).

We specify the action and reward.
Depending on the six actions of the agent, $a = 0, 1, \cdots, 5$, the
parameters in Eq.~(\ref{2DGauss}) are changed in each time step as
\begin{equation}
  \begin{array}{cc}
    a = 0 & \xi \rightarrow \xi + 0.15, \\
    a = 1 & \xi \rightarrow \xi - 0.15, \\
    a = 2 & \eta \rightarrow \eta + 0.15, \\
    a = 3 & \eta \rightarrow \eta - 0.15, \\
    a = 4 & A_\perp \rightarrow A_\perp + 2, \\
    a = 5 & A_\perp \rightarrow A_\perp - 2,
  \end{array}
\end{equation}
where the displacement 0.15 is identical with the numerical mesh size.
Since the strength $A_\perp$ of the Gaussian potential produced by a
blue-detuned laser beam should be nonnegative, the actions $a = 5$ is
ignored when $A_\perp$ becomes negative.
The fidelity between the wave function $\psi_\perp(x, y, t)$ at time $t$ and
the target wave function $\psi_{\rm target}(x, y)$ is defined by
\begin{equation} \label{F}
  F(t) = \left| \int \psi_{\rm target}^*(x, y) \psi_\perp(x, y, t) dxdy
  \right|^2.
\end{equation}
Since the present purpose is to increase the fidelity as much as possible,
the reward $r_t$ given to the agent should be taken to be a monotonically
increasing function of the fidelity.
Here, to enhance the increase of the fidelity near $F = 1$, we take
the reward as
\begin{equation} \label{reward16}
  r_t = F(t) + 8 [F(t)]^{16}.
\end{equation}
As shown in the inset of Fig.~\ref{f:ev2d}(g), this function steeply rises
for $F \gtrsim 0.9$.

We performed the reinforcement learning in Algorithm 1 for $N_{\rm episode}
= 5000$ episodes.
The total reward obtained in each episode is defined as
\begin{equation} \label{Rtot}
  R_{\rm total} = \sum_{n = 0}^{500} r_{n \Delta t},
\end{equation}
where the episode terminates at $T_{\rm end} = 500 \Delta t$.
The episode with the largest $R_{\rm total}$ among 5000 episodes is shown in
Fig.~\ref{f:ev2d}.
First, the Gaussian potential is moved leftward (Fig.~\ref{f:ev2d}(b)),
and at the edge of the BEC a clockwise vortex is released to the periphery
of the BEC, leaving a counterclockwise vortex in the Gaussian potential
(Fig.~\ref{f:ev2d}(c)).
The Gaussian potential then carries the counterclockwise vortex to the
center of the BEC (Fig.~\ref{f:ev2d}(d)), at which point the strength of the
Gaussian potential is reduced to produce the desired state
(Fig.~\ref{f:ev2d}(e)).
This process finishes at $t \simeq 130$, and after that the vortex is kept
at the center.
It is numerically predicted~\cite{Frisch} and experimentally
verified~\cite{Neely} that the simple translation of a circular potential in
a BEC produces a pair of clockwise and counterclockwise vortices
simultaneously.
In the above dynamics, in contrast, the strong anisotropy at the edge of the
condensate is used to release only a single vortex from the potential.

Figure~\ref{f:ev2d}(g) shows the time dependence of the controlled
parameters $\xi(t)$, $\eta(t)$, and $A(t)$.
The strength $A(t)$ of the Gaussian potential is first decreased from the
initial value as the potential goes to the edge of the BEC.
This is because the density is low at the edge of the BEC and a weak
Gaussian potential is suitable for manipulating the vortices.
The strength $A(t)$ is then increased to rapidly transport the
counterclockwise vortex to the center.
Even after the final state is produced at $t \simeq 130$, $A(t)$ is kept at
$\simeq 4$, which helps fix the vortex to the center.
It was confirmed that the vortex stays almost at the center, even if $A(t)$
vanishes for $t \geq 130$.

\begin{figure}[tb]
\begin{center}
  \includegraphics[width=8cm]{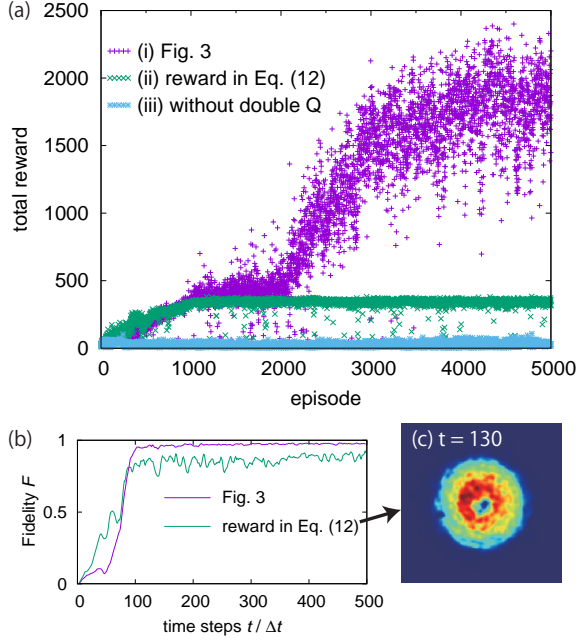}
\end{center}
\caption{
  (a) Total reward in Eq.~(\ref{Rtot}) obtained in each episode in the
  learning process.
  (i) The learning process with 5000 episodes.
  The episode with the largest total reward is used in Fig.~\ref{f:ev2d}.
  (ii) Using the reward in Eq.~(\ref{linear}) instead of
  Eq.~(\ref{reward16}).
  (iii) Using the target value in Eq.~(\ref{nodouble}) instead of the
  double-Q learning in Eq.~(\ref{yt}).
  (b) Time evolution of the fidelity for the episode with the best total
  reward for conditions (i) and (ii).
  (c) Snapshot of the density profile for case (ii) in (b).
}
\label{f:reward2d}
\end{figure}
Figure~\ref{f:reward2d}(a) shows the total reward $R_{\rm total}$ obtained
in each episode.
The total reward increases as the agent experiences more episodes, and
finally reaches $R_{\rm total} \simeq 2000$.
The total reward first increases to $R_{\rm total} \simeq 400$, then remains
on a plateau till the $\simeq 2000$th episodes and then increases again.
This behavior is related to the form of the reward in the inset in
Fig.~\ref{f:ev2d}(g), because the total reward saturates at $R_{\rm total}
\simeq 400$ if we replace the reward in Eq.~(\ref{reward16}) with
\begin{equation} \label{linear}
r_t = F(t).
\end{equation}
Figure~\ref{f:reward2d}(b) shows the time evolution of the fidelity $F(t)$
for the best episode, which indicates that the fidelity reaches $F \simeq
0.97$ using the reward in Eq.~(\ref{reward16}), while $F \simeq 0.9$ for the
reward in Eq.~(\ref{linear}).
A snapshot of the density profile for the reward in Eq.~(\ref{linear})
is shown in Fig.~\ref{f:reward2d}(c), which exhibits more
short-wavelength excitations than Fig.~\ref{f:ev2d}(e).
Thus, the nonlinear term in Eq.~(\ref{reward16}) makes the fidelity closer
to unity, and reduces short-wavelength excitations.
In Fig.~\ref{f:reward2d}(a), we also show the case without double-Q
learning, i.e., we use
\begin{equation} \label{nodouble}
  y_t = r_t + \gamma Q(s_{t+\Delta t}, \mbox{argmax}_a Q(s_{t+\Delta t},
  a))
\end{equation}
instead of Eq.~(\ref{yt}).
Without the target network $\hat Q$, the learning process does not proceed,
which indicates the validity of the target network.

\subsection{Three-dimensional system}

\begin{figure}[tb]
\begin{center}
\includegraphics[width=7cm]{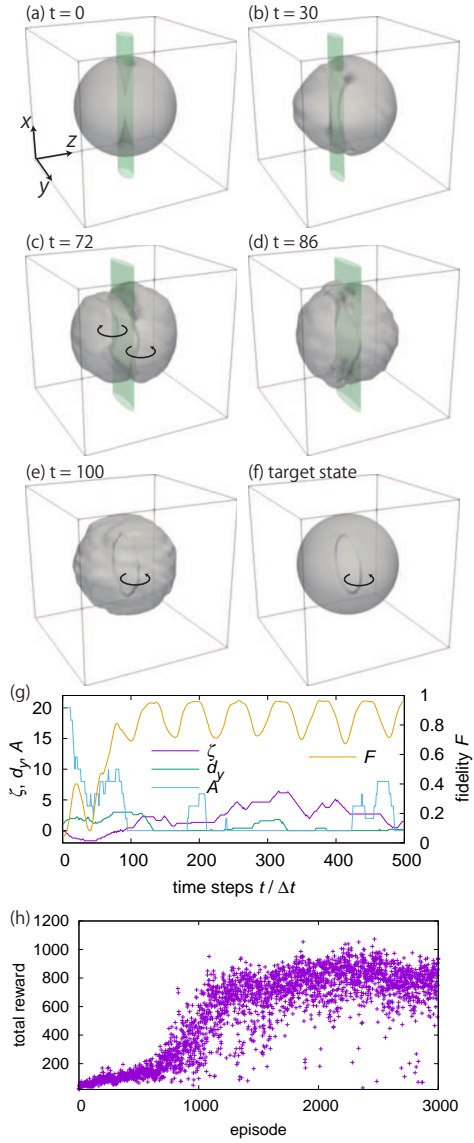}
\end{center}
\caption{
  (a)-(e) Dynamics of a 3D BEC with the best total reward.
  The BEC is confined in an isotropic harmonic trap with an external
  Gaussian potential in Eq.~(\ref{3DGauss}).
  (f) Target state containing a stationary vortex ring.
  The arrows indicate the directions of circulation.
  Surfaces of equal density $|\psi|^2 = 0.0005$ are shown in gray and
  surfaces of equal Gaussian potential $V_G = 3$ are shown in green (dark
  gray).
  The size of the cubic frames is $16 \times 16 \times 16$.
  (g) Parameters $\zeta(t)$, $d_y(t)$, and $A(t)$ of the Gaussian
  external potential used in (a)-(e).
  For $t / \Delta t \gtrsim 100$, the Gaussian potential almost vanishes,
  since $A(t) \simeq 0$ or $d_y(t) \simeq 0$.
  The time evolution of the fidelity $F(t)$ defined in Eq.~(\ref{F3D}) is
  also shown.
  (h) Total reward obtained in each episode in the learning process.
  The episode with the best total reward among these 3000 episodes is used
  in (a)-(e) and (g).
  See the Supplemental Material for a movie of the dynamics shown in
  (a)-(e)~\cite{movie3D}.
}
\label{f:ev3d}
\end{figure}
Next, we consider a 3D system obeying Eq.~(\ref{GP}), in which a BEC is
confined in an isotropic harmonic potential, and a Gaussian laser beam is
applied from the $x$ direction, producing the potential in
Eq.~(\ref{3DGauss}).
The initial parameters in Eq.~(\ref{3DGauss}) are set to $A(0) = 20$,
$d_y(0) = 1$, and $\zeta(0) = 0$, and $d_z = 0.5$ is fixed.
The initial wave function is the ground state for these parameters, as shown
in Fig.~\ref{f:ev3d}(a).
Here, our target state $\psi_{\rm target}(\bm{r})$ is the stationary state
containing a vortex ring~\cite{Crasovan} with axial symmetry about the $z$
axis, as shown in Fig.~\ref{f:ev3d}(f).
Although a vortex ring in a uniform system travels in the direction of the
symmetry axis, the vortex ring in Fig.~\ref{f:ev3d}(f) is stationary due to
the inhomogeneity.
Numerically, this target state is generated by imprinting the phase
$e^{i \phi}$ with $\phi = \tan^{-1}[z / (r_\perp - r_0)] - \tan^{-1}[z /
  (r_\perp + r_0)]$ followed by the imaginary-time evolution, where $r_0$
and the duration of the imaginary-time evolution are chosen appropriately.

As in the 2D case, the agent chooses one of six actions, $a = 0, 1, \cdots,
5$, according to which the parameters of the Gaussian potential are modified
as
\begin{equation}
  \begin{array}{cc}
    a = 0 & \zeta \rightarrow \zeta + 0.15, \\
    a = 1 & \zeta \rightarrow \zeta - 0.15, \\
    a = 2 & d_y \rightarrow d_y + 0.2, \\
    a = 3 & d_y \rightarrow d_y - 0.2, \\
    a = 4 & A \rightarrow A + 2, \\
    a = 5 & A \rightarrow A - 2,
  \end{array}
\end{equation}
in each time step.
The actions $a = 3$ and 5 are ignored when $d_y$ and $A$ become negative,
respectively.
The reward has the same form as in Eq.~(\ref{reward16}) with the fidelity
being given by
\begin{equation} \label{F3D}
F(t) = \left| \int \psi^*_{\rm target}(\bm{r}) \psi(\bm{r}, t) d\bm{r}
\right|^2.
\end{equation}

Figures~\ref{f:ev3d}(a)-\ref{f:ev3d}(e) show the time evolution of the BEC
for the episode with the best total reward among $N_{\rm episode} = 3000$
episodes.
The controlled parameters of the Gaussian external potential are shown in
Fig.~\ref{f:ev3d}(g).
First, the Gaussian potential is moved in the $-z$ direction, and a pair of
vortex-antivortex lines are produced (Fig.~\ref{f:ev3d}(b)).
The width $d_y$ of the Gaussian potential is then increased
(Fig.~\ref{f:ev3d}(c)), and the two vortex lines connect with each other at
the $\pm x$ edges (Fig.~\ref{f:ev3d}(d)), which forms a vortex ring
(Fig.~\ref{f:ev3d}(e)).
The produced vortex ring is almost stationary and stays in the condensate
until the end of the episode $t / \Delta t = 500$.
After $t / \Delta t \gtrsim 100$, the Gaussian external potential almost
vanishes because either $A = 0$ or $d_y = 0$.

The generation of the vortex ring shown above is nontrivial, because the
external Gaussian potential is uniform in the $x$ direction.
In a uniform system, using such a potential, we can only generate vortex
lines along the $x$ direction, and therefore the vortex-ring formation in
our system is due to the inhomogeneity in the trap.
The board-like potential with large $d_y$, as in Figs.~\ref{f:ev3d}(c) and
\ref{f:ev3d}(d), also plays an important role.
This potential creates the constrictions at the $\pm x$ edges of the
isodensity surface, as shown in Fig.~\ref{f:ev3d}(c), which bend the vortex
lines and forms a ring.

The time evolution of the fidelity $F(t)$ is shown in Fig.~\ref{f:ev3d}(g).
Unlike the 2D case in Fig.~\ref{f:reward2d}(b), the fidelity in
Fig.~\ref{f:ev3d}(g) exhibits nonadiabatic oscillation, which may be
unavoidable for the present condition.
Figure~\ref{f:ev3d}(h) shows the total reward $R_{\rm total}$ obtained in
each episode.
The total reward first increases gradually for $\lesssim 1000$th step and
then suddenly rises to $R_{\rm total} \simeq 800$.
This behavior is similar to that in the 2D case shown in
Fig.~\ref{f:reward2d}(a).

\section{Conclusions and discussions}
\label{s:conc}

We have applied reinforcement learning to the control of nonlinear matter
waves.
The agent in the reinforcement learning is implemented using a deep
convolutional neural network (CNN), and the state of the Bose-Einstein
condensate (BEC) is input into the CNN to determine the next action of the
agent.
According to the action of the agent, the position and shape of the external
Gaussian potential applied to the BEC are controlled. 
The agent is trained so that the state of the BEC approaches the prescribed
target state.

Using this method, we demonstrated that quantized vortices can be created
and manipulated in 2D and 3D systems.
In the 2D system, the target state was set to be the single-vortex state at
the center.
Although vortices and antivortices are always created in pairs, the agent
found a way to expel one of them to leave only a single vortex at the center
of the BEC (Fig.~\ref{f:ev2d}).
In the 3D system, the target state was set to be the stationary vortex-ring
state.
Although the Gaussian potential is 2D (uniform in the $x$ direction), the
agent found a way to produce such a 3D structure (Fig.~\ref{f:ev3d}).

The two examples demonstrated in the present paper are interesting, but
rather simple;
the control parameters found in Figs.~\ref{f:ev2d}(g) and \ref{f:ev3d}(g)
for creating the target states are very simple functions.
Other methods have been developed to control quantum systems~\cite{Khaneja,
  Doria, Caneva} that may find similar optimal results.
Furthermore, the simple vortex states used for our target states can also be
created by other methods, such as methods using Rabi
transitions~\cite{Anderson,Ruo}.
In this sense, our method might be considered overcomplicated for the
present examples.
However, reinforcement learning is very versatile and may prove to be an
effective approach for handling more complicated problems (e.g.,
manipulation of multiple vortex states, control of quantum turbulence,
creation of nontrivial topological excitations, etc.).
This will be the topic of future studies.

Another challenging extension of this study will be the real-time control of
BECs in experiments, where a series of nondestructive imaging data is given
to the agent, and the reward is based on the measurement data.
Reinforcement learning may be suitable for such measurement-based control
which introduces errors and noise, since the method has already been
successfully applied to control real-world objects, such as robots.

\begin{acknowledgments}
This research was supported by JSPS KAKENHI Grant Numbers JP17K05595 and
JP17K05596.
\end{acknowledgments}

\end{document}